# scientific reports

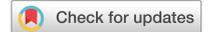

OPEN

# Variations in climate habitability parameters and their effect on Earth's biosphere during the Phanerozoic Eon

Iva Vilović[1✉], Dirk Schulze-Makuch[1,2,3,4] & René Heller[5,6]

Essential insights on the characterization and quality of a detectable biosphere are gained by analyzing the effects of its environmental parameters. We compiled environmental and biological properties of the Phanerozoic Eon from various published data sets and conducted a correlation analysis to assess variations in parameters relevant to the habitability of Earth's biosphere. We showed that environmental parameters such as oxygen, global average surface temperatures, runoff rates and carbon dioxide are interrelated and play a key role in the changes of biomass and biodiversity. We showed that there were several periods with a highly thriving biosphere, with one even surpassing present day biodiversity and biomass. Those periods were characterized by increased oxygen levels and global runoff rates, as well as moderate global average surface temperatures, as long as no large or rapid positive and/or negative temperature excursions occurred. High oxygen contents are diagnostic of biomass production by continental plant life. We find that exceptionally high oxygen levels can at least in one instance compensate for decreased relative humidities, providing an even more habitable environment compared to today. Beyond Earth, these results will help us to understand how environmental parameters affect biospheres on extrasolar planets and guide us in our search for extraterrestrial life.

Habitability has been defined as the ability of an environment to support the activity of at least one known organism[1]. Environmental factors, such as temperature, oxygen content, relative humidity and carbon dioxide levels have played a significant role in the ecology of our biosphere throughout geological time[2]. Specifically, during the Phanerozoic Eon, which spans from 542 Ma ago until today, habitable ecological regions fragmented into niches and complex life arose. We use the amount of biomass production and biodiversity as a first measure of the habitability of Earth's biosphere during the Phanerozoic. Variations in habitability are exhibited during some of the extensive "greenhouse" phases in the last 541 Ma, during which biodiversity has remained relatively low and was accompanied by relatively high extinction rates[3,4]. Previous work focused on how global average surface temperatures ($T_{glob}$) affect the biosphere[2,5], which are strongly linked to the amount of atmospheric carbon dioxide levels ($CO_2$)[6]. Large injections of $CO_2$ into the atmosphere (e.g. via volcanic eruptions[7]) lead to rapid increases of $T_{glob}$, whereas a deficit in $CO_2$ levels is commonly connected to glaciation periods[6]. $CO_2$ levels, in turn, are influenced by the amount of atmospheric oxygen partial pressures ($pO_2$), which are also of great significance for biomass and biodiversity[8–16]. For example, ca. 325–275 Ma ago, when the oxygen levels on Earth were higher than today, enough energy was available for giant insects with wingspans of 70 cm to inhabit Earth[17]. It is also hypothesized that prior to this time, i.e. around 372–358 Ma ago, the decline in available marine oxygen levels was one of the conditions leading up to the late Devonian extinction event[18,19]. The biodiversity of a marine ecosystem is also strongly affected by global runoff rates (RR), which can be used to infer the atmospheric humidity or aridity of certain geological time periods[20]. Higher RR are often connected to the largest input of macronutrients into coastal marine environments, providing more energetics for growth and production in the marine biosphere[21–25]. RR exhibits a concomitant relationship with Earth's varying continental configurations, a projection of modern day latitudinal rainfall and rates of evaporation[26–30]. An example of the influence of high and low RR on biodiversity today can be seen in the tropical rainforests and the Atacama Desert, respectively[31–33].

[1]Astrobiology Research Group, Zentrum für Astronomie und Astrophysik, Technische Universität Berlin, 10623 Berlin, Germany. [2]GFZ German Research Center for Geosciences, Section Geomicrobiology, 14473 Potsdam, Germany. [3]Department of Experimental Limnology, Leibniz-Institute of Freshwater Ecology and Inland Fisheries (IGB), 16775 Stechlin, Germany. [4]School of the Environment, Washington State University, Pullman, Washington, USA. [5]Max-Planck-Institut für Sonnensystemforschung, 37077 Göttingen, Germany. [6]Institut für Astrophysik, Georg-August-Universität Göttingen, 37077 Göttingen, Germany. ✉email: iva.vilovic@campus.tu-berlin.de

Here we use previously published data sets from 5 studies and 7 databases[27,34–40] (see Methods section) of the environmental parameters $pO_2$, $T_{glob}$, RR and $CO_2$ to gain more insights on the extent of their impact on biomass production and biological diversity. Our hypothesis is that these input parameters play a deciding role in the changes of biomass and biodiversity and the overall habitability of Earth, ultimately also impacting its detectability. The biodiversity data[34] in this work is based on marine fauna, which is considered a good proxy for overall biodiversity because the majority (~ 75%) of the faunal (i.e. animal) biomass today is of marine origin[41].

## Results

We performed a cross-correlation analysis in order to quantify the similarities between the environmental variables (i.e. $pO_2$, $T_{glob}$, RR and $CO_2$) and the biological parameters (i.e. biomass, and biodiversity) as well as between the environmental variables themselves. The cross-correlation function (ccf) provides a correlation value of two time series at a temporal lag of 0 ($ccf_0$), as well as determines at which temporal lag ($\mu^*$) the time series exhibit a maximum correlation. A numerical synthesis of the results is presented in Table 1.

Panels (a) and (e) of Fig. 1 depict the correlation of the mean atmospheric $pO_2$ curve (based on three data sets, see Fig. 4 in the Methods section) with biomass and biodiversity, respectively. The mean $pO_2$ exhibits a cyclical nature in the early Phanerozoic, reaching its maximum above the present atmospheric level (PAL) during the transition from the mid-Carboniferous to early Triassic period (~ 325–240 Ma ago). It stays relatively constant with a small decline in the last 140 Ma. The mean $pO_2$ curve exhibits a strong positive correlation with biomass for the entire Phanerozoic ($ccf_0 = 3.99 \times 10^{-3}$). When separately considering the two temporal ranges, the correlation between $pO_2$ and biomass is even stronger before 300 Ma ago ($ccf_0 = 1.00 \times 10^{-2}$) and still substantial thereafter ($ccf_0 = 6.54 \times 10^{-3}$). The maximum correlation between $pO_2$ and biomass for the entire timespan as well as for the most recent 300 Ma is reached when $pO_2$ is shifted backwards by 4 Ma, indicating that biomass precedes (i.e. leads) $pO_2$ in those temporal ranges (Table 1, Fig. 2 in main document and S1 and S2 in Supplementary Materials). For the time periods before 300 Ma, $pO_2$ and biomass already exhibit a maximum correlation at a lag of 0, indicating a concurrent trend in their evolutions.

| Environmental variables | | | | | | | | | | |
|---|---|---|---|---|---|---|---|---|---|---|
| | | | $pO_2$* | | $T_{glob}$[35] | | RR[42] | | $CO_2$[40] | |
| | | | $ccf_0$ | $\mu^* \pm 1\sigma$ [Ma] | $ccf_0$ | $\mu^* \pm 1\sigma$ [Ma] | $ccf_0$ | $\mu^* \pm 1\sigma$ [Ma] | $ccf_0$ | $\mu^* \pm 1\sigma$ [Ma] |
| Biological variables | Biomass[36] | Total | **3.99 × 10⁻³** | **−4 ± 40/36** | − 1.49 × 10⁻³ | 16 ± 19/31 | − 2.31 × 10⁻³ | −24 ± 54/47 | − 4.14 × 10⁻³ | −14 ± 61/57 |
| | | Before 300 | **1.00 × 10⁻²** | **0 ± 26/27** | − 1.12 × 10⁻² | 8 ± 12/11 | − 6.02 × 10⁻³ | −24 ± 69/31 | − 1.77 × 10⁻² | 0 ± 28/29 |
| | | After 300 | **6.54 × 10⁻³** | **−4 ± 22/22** | − 5.90 × 10⁻³ | 14 ± 21/30 | **6.46 × 10⁻³** | 10 ± 27/28 | **4.84 × 10⁻³** | **8 ± 30/26** |
| | Bio-diversity[34] | Total | − 2.24 × 10⁻³ | 24 ± 50/92 | − 4.22 × 10⁻³ | 2 ± 26/50 | **3.98 × 10⁻³** | 8 ± 80/50 | **3.91 × 10⁻³** | 10 ± 57/48 |
| | | Before 300 | − 1.02 × 10⁻² | −18 ± 48/30 | 1.49 × 10⁻³ | −24 ± 36/22 | **1.00 × 10⁻²** | 0 ± 26/25 | **9.33 × 10⁻³** | 14 ± 40/35 |
| | | After 300 | **4.19 × 10⁻³** | **−10 ± 25/36** | − 1.17 × 10⁻² | 2 ± 15/17 | **6.18 × 10⁻³** | 16 ± 37/32 | 1.55 × 10⁻³ | − 6 ± 27/24 |
| Environmental variables | $pO_2$* | Total | **3.64 × 10⁻³** | **0 ± 59/59** | | | | | | |
| | | Before 300 | **8.00 × 10⁻³** | **0 ± 43/43** | | | | | | |
| | | After 300 | **6.67 × 10⁻³** | **0 ± 23/23** | | | | | | |
| | $T_{glob}$[35] | Total | − 3.03 × 10⁻⁴ | −24 ± 24/52 | **3.64 × 10⁻³** | **0 ± 18/18** | | | | |
| | | Before 300 | − 7.12 × 10⁻³ | −10 ± 60/57 | **8.00 × 10⁻³** | **0 ± 16/16** | | | | |
| | | After 300 | − 2.25 × 10⁻³ | −16 ± 21/32 | **6.67 × 10⁻³** | **0 ± 12/12** | | | | |
| | RR[42] | Total | − 4.21 × 10⁻³ | 24 ± 56/104 | − 2.21 × 10⁻³ | 24 ± 36/90 | **3.64 × 10⁻³** | **0 ± 60/60** | | |
| | | Before 300 | − 7.40 × 10⁻⁴ | 24 ± 38/30 | − 2.03 × 10⁻³ | 24 ± 16/22 | **8.00 × 10⁻³** | **0 ± 25/25** | | |
| | | After 300 | **5.00 × 10⁻³** | −8 ± 19/19 | − 1.10 × 10⁻³ | 24 ± 36/102 | **6.67 × 10⁻³** | **0 ± 28/28** | | |
| | $CO_2$[40] | Total | − 7.79 × 10⁻³ | 0 ± 70/70 | 1.33 × 10⁻³ | −18 ± 14/15 | **3.30 × 10⁻³** | −24 ± 134/71 | **3.64 × 10⁻³** | **0 ± 81/81** |
| | | Before 300 | − 1.58 × 10⁻² | 0 ± 44/45 | **8.00 × 10⁻³** | **0 ± 95/40** | 2.20 × 10⁻³ | | **8.00 × 10⁻³** | **0 ± 40/0** |
| | | After 300 | **6.32 × 10⁻³** | −8 ± 23/23 | 3.08 × 10⁻³ | −22 ± 16/16 | **6.33 × 10⁻³** | −24 ± 34/60<br>−12 ± 22/22 | **6.67 × 10⁻³** | **0 ± 17/17** |

**Table 1.** Summary of cross-correlation analysis between environmental parameters (i.e. mean atmospheric oxygen partial pressures $pO_2$, global average surface temperatures $T_{glob}$, global runoff rates RR, and atmospheric carbon dioxide levels $CO_2$) and biological variables (i.e. total biodiversity of marine fauna and biomass). $ccf_0$ is the magnitude of the cross-correlation function (ccf) between two time series at a temporal lag of 0. $\mu^*$ is the temporal lag at which the time series exhibit a maximum correlation. Negative $\mu^*$ indicate that a column variable precedes (i.e. leads) the corresponding row variable and vice versa. The abbreviations "before 300" and "after 300" refer to the temporal regions before 300 Ma ago, and after 300 Ma ago, respectively. *Based on the mean and standard deviation calculations of three oxygen curves (see Fig. 4 and Table 3). Significant values are in [bold].

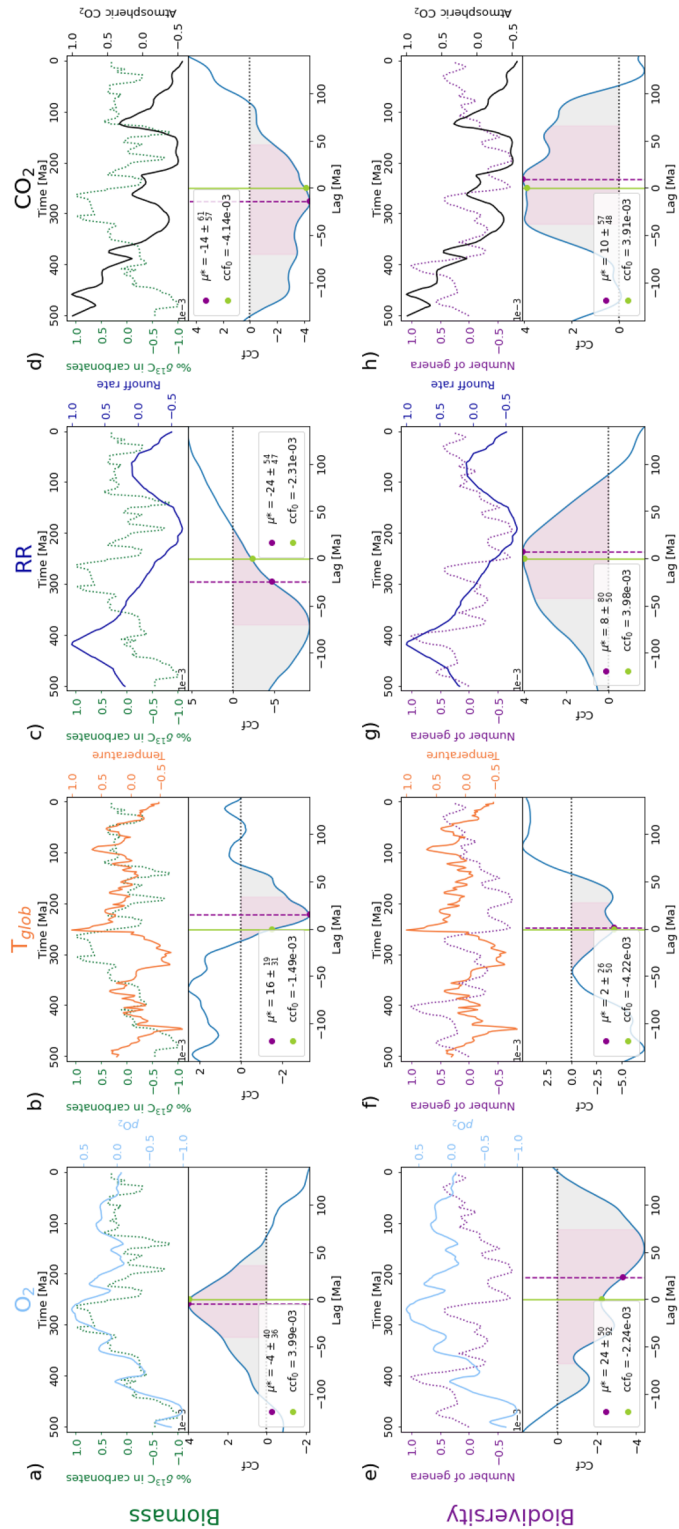

**Figure 1.** Environmental and biological parameters through Earth history from 500 Ma before present until today and their corresponding cross-correlation functions (ccf). The ccf provides a correlation value of the two time series at a temporal lag of 0 ($ccf_0$), as well as determines at which temporal lag ($\mu^*$) the time series exhibit a maximum correlation. The environmental parameters include in the upper panels of (**a**) through (**h**), respectively: mean atmospheric oxygen partial pressures $pO_2$, global average surface temperatures $T_{glob}$ runoff rates RR, and atmospheric carbon dioxide levels $CO_2$. The biological parameters include in the upper panels of (**a**) through (**d**) the primary productivity (in ‰ $\delta^{13}C$ in carbonates) as a proxy for biomass (dashed dark green curves) and in the upper panels of (**e**) through (**h**) the biodiversity (number of genera) of marine fauna (dotted magenta curves). Each parameter was normalized to its maximum (absolute) value after the subtraction of its corresponding mean value. This way each time series fluctuates around 0 and has the same normalized units. Bottom panels of (**a**) through (**h**): Corresponding cross-correlation function (ccf). We constrained our analysis to a ± 25 Ma lag since ever larger temporal ranges have a decreasing biogeochemical significance. The positive and negative errors with respect to $\mu^*$ were calculated by determining the positive/negative temporal lags at which 34.15% of the total area between the ccf and the horizontal zero-line is encapsulated.

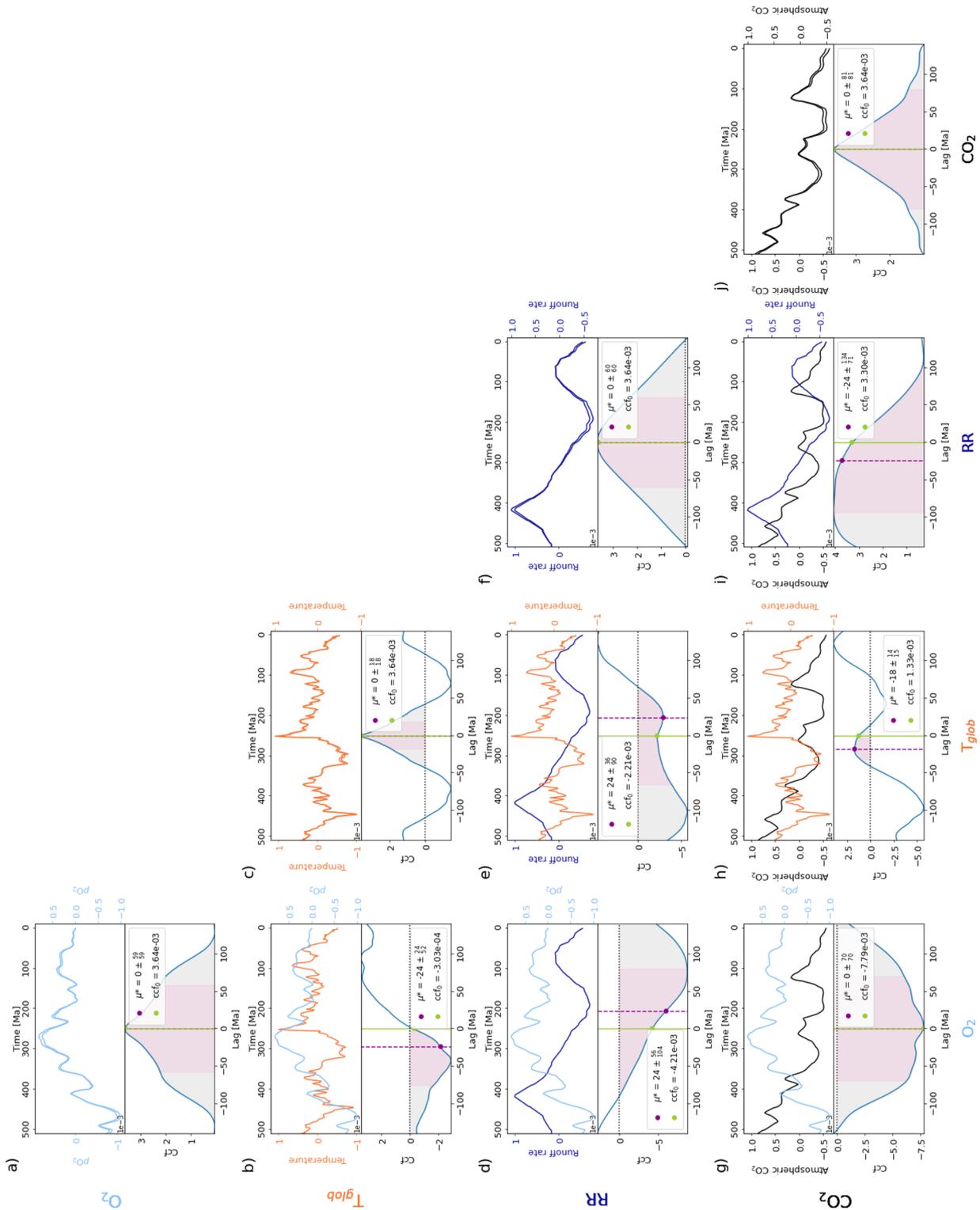

**Figure 2.** Environmental parameters through Earth history from 500 Ma before present until today and their corresponding cross-correlation functions (ccf). They include from top to bottom and left to right panel respectively: mean atmospheric oxygen partial pressures $pO_2$, global average surface temperatures $T_{glob}$, runoff rates RR, and atmospheric carbon dioxide $CO_2$. The auto-correlations are located on the diagonal. Other markers as in Fig. 1.

The mean $pO_2$ and biodiversity data exhibit a weakly negative correlation in total ($ccf_0 = -2.24 \times 10^{-3}$), which exhibit a strong anticorrelation before 300 Ma ($ccf_0 = -1.02 \times 10^{-2}$) but a strong positive correlation from then on ($ccf_0 = 4.19 \times 10^{-3}$). A maximum anticorrelation between $pO_2$ and biodiversity before 300 Ma is reached when $pO_2$ is shifted backwards by 18 Ma, whereas the maximum correlation after 300 Ma is reached when $pO_2$ is shifted backwards by 10 Ma, indicating for both geological time frames that biodiversity precedes $pO_2$ (Table 1, Fig. 2 in main document and S1 and S2 in Supplementary Materials). Furthermore, $pO_2$ correlates negatively with varying degrees of magnitude with all environmental parameters, except for RR and $CO_2$ in the most recent 300 Ma ($ccf_0 = 5.00 \times 10^{-3}$ and $ccf_0 = 6.32 \times 10^{-3}$, respectively) (Table 1, Fig. 2 and S3 and S4 in Supplementary Materials).

The correlation of $T_{glob}$ with biomass and biodiversity throughout the Phanerozoic is shown in Fig. 1b and f, respectively. The curve of $T_{glob}$ alternates between warm and cool periods throughout this era. The cool periods, during which $T_{glob}$ was below 15 °C, included the mid Neogene to present (10–0 Ma ago), the transitional period between the Carboniferous and Permian (275–325 Ma ago) and the Ordovician–Silurian transitional period (430–460 Ma ago). Times of higher $T_{glob}$ relative to today include the Cambrian (542–485 Ma ago), the late Devonian/ early Carboniferous (~375–350 Ma ago) and the mid Cretaceous to early Paleogene periods (~100–50 Ma ago). Overall, the $T_{glob}$ curve exhibits a weak negative correlation with biomass with $ccf_0 = -1.49 \times 10^{-3}$. When analyzing their similarities separately for the geological time periods before and after 300 Ma, they exhibit an even stronger anticorrelation with $ccf_0 = -1.12 \times 10^{-2}$ in the earlier periods than in the second range from 300 Ma to today. This can be observed towards the early Ordovician period (~500–480 Ma ago) when biomass reached a global minimum and temperatures a local maximum, and the late Carboniferous/early Permian period (~340–260 Ma ago) during which biomass reached a global maximum but was accompanied by one of the largest drops in Phanerozoic surface temperatures. The maximum anticorrelation between $T_{glob}$ and biomass for the entire timespan is reached when $T_{glob}$ is shifted forward by 16 Ma, indicating that $T_{glob}$ precedes (i.e. leads) biomass (Table 1, Fig. 2). For the time periods before and after 300 Ma, $T_{glob}$ precedes biomass by 8 and 14 Ma, respectively (Table 1, S1 and S2 in Supplementary Materials).

Temperature and biodiversity (Fig. 1f) correlate moderately negatively with $ccf_0 = -4.22 \times 10^{-3}$ during the entire timespan of the Phanerozoic. At times of temperature extrema, both positive and negative, the biodiversity exhibits local minima, e.g. around the end of the Ordovician period ~425–450 Ma ago and during the Jurassic/Triassic periods between 250 and 150 Ma ago, which correlate with the Late Ordovician, the Permian–Triassic and the Triassic-Jurassic extinction events, respectively. This effect is enhanced if periods of high temperature are accompanied by lower RRs (see Fig. 1g). The maximum anticorrelation between $T_{glob}$ and biodiversity for the entire timespan as well as for the most recent 300 Ma is reached when $T_{glob}$ is shifted forwards by 2 Ma, indicating that $T_{glob}$ is leading biodiversity in those temporal ranges (Table 1, Fig. 2 in main text and S2 in Supplementary Materials). In general, $T_{glob}$ mostly correlates negatively with other environmental parameters such as $pO_2$ and RR, but positively with $CO_2$ throughout the Phanerozoic (Table 1, Fig. 2 and S3 and S4 in Supplementary Materials).

The correlation of the global runoff rates with biomass and biodiversity is shown in Fig. 1c and g, respectively. The RR curve (Fig. 1c and g) indicates highest values during the Cretaceous (~100–65 Ma ago), the Devonian-Silurian (~450–400 Ma ago), and the Cambrian (~550–485 Ma ago) periods, whereas the lowest RR occurred during the transitional period between the Triassic and Jurassic (~250–150 Ma ago) and the Neogene-Holocene (~20–0 Ma ago). The results exhibit a strong positive correlation ($ccf_0 = 6.46 \times 10^{-3}$) between RR and biomass for the most recent 300 Ma, while there is an equal in magnitude negative correlation for the earlier time period (Table 1). In the most recent 300 Ma, a maximum positive correlation between RR and biomass is reached when RR is shifted forward in time by 16 Ma, indicating that RR leads biomass for that temporal range.

The results overall show a strong positive correlation of $ccf_0 = 3.98 \times 10^{-3}$ between RR and biodiversity. This correlation reaches a maximum when RR is shifted forwards in time by 8 Ma, indicating that RR precedes biodiversity throughout the Phanerozoic. The correlation is stronger for the more recent time period (300 Ma—today) with an even higher degree of magnitude for the time period before 300 Ma during which RR and biodiversity already exhibit a maximum correlation at a lag of 0, indicating a concurrent evolution trend (Table 1, Fig. 2 in main text and S2 in Supplementary Materials). RR also correlates negatively with environmental parameters such as $T_{glob}$ and $pO_2$, yet positively with $CO_2$ throughout the Phanerozoic and with $pO_2$ in the most recent 300 Ma (Table 1, Fig. 2 and S3 and S4 in Supplementary Materials).

The correlation of $CO_2$ with biomass and biodiversity throughout the Phanerozoic is shown in Fig. 1d and h, respectively. The curve of $CO_2$ exhibits an oscillating nature with a generally decreasing trend towards the present. The periods with similar levels of $CO_2$ include the Jurassic (~200–150 Ma ago) and the mid to late Carboniferous (~320–310 Ma ago). Other periods during the Phanerozoic were marked by higher levels compared to the present.

Overall, the $CO_2$ curve exhibits a negative correlation with biomass with $ccf_0 = -4.14 \times 10^{-3}$ which further increases when analyzing the temporal region before 300 Ma ago ($ccf_0 = -1.77 \times 10^{-2}$). A maximum anticorrelation for the entire Phanerozoic temporal range is reached when $CO_2$ is shifted backwards by 14 Ma, indicating that biomass precedes $CO_2$, while a maximum anticorrelation between $CO_2$ and biomass before 300 Ma ago is already reached without a temporal shift. Contrastingly, the results show a positive correlation between $CO_2$ and biomass in the most recent 300 Ma ($ccf_0 = 4.84 \times 10^{-3}$), which reaches a maximum when $CO_2$ is shifted forwards by 8 Ma. This indicates that $CO_2$ leads biomass for that temporal range (Table 1, Fig. 2 in main text and S1 and S2 in Supplementary Materials). $CO_2$ and biodiversity (Fig. 1h,) correlate strongly positively with $ccf_0 = 3.91 \times 10^{-3}$ for the entire timespan of the Phanerozoic with an even stronger positive correlation exhibited before 300 Ma ago ($ccf_0 = 9.33 \times 10^{-3}$). A maximum correlation for the entirety of the Phanerozoic is reached when $CO_2$ is shifted forwards by 10 Ma, indicating a preceding relationship with biodiversity (Table 1, Fig. 2 in main text and S1 and S2 in Supplementary Materials). In general, $CO_2$ mostly correlates positively throughout the Phanerozoic with other environmental parameters such as RR and $T_{glob}$, but negatively with $pO_2$ except for in the most recent 300 Ma (Table 1 and Fig. 2 in main text and S3 and S4 in Supplementary Materials).

### Discussion
We assessed the variations of Earth's global biosphere during the Phanerozoic eon based on the effects that atmospheric oxygen partial pressures ($pO_2$), global average surface temperatures ($T_{glob}$), runoff rates (RR) and atmospheric carbon dioxide levels ($CO_2$) had on biomass and biodiversity. This can be used as a stepping stone for the characterization of potential biospheres on extrasolar worlds, which necessarily influence the quantitative measures of habitability in our search for life elsewhere. Caution is required to interpret the data, because the correlations obtained in this work are not clear-cut and there are likely other factors not included in the analysis affecting the environmental parameters as well as the biology of our planet.

However, having said that, certain take-away messages seem to be evident (Fig. 3, Table 2). $pO_2$ seems to always be strongly positively correlated with biomass. A high atmospheric $pO_2$ is indicative of photosynthesis. Atmospheric $pO_2$ can also be used for primary production (i.e. biomass) via some autotrophic organisms that use it for aerobic cellular respiration. To be noted, though, is that excessive oxygen and carbon dioxide levels can also inhibit photosynthesis[43]. The results indicate that there was a concurrent evolution of biomass and oxygen prior to the spreading of land plants around 300 Ma ago, after which biomass started leading (i.e. preceding) the production of oxygen by up to 4 Ma (Table 1). This indicates that the dominant organisms (i.e. oxygenic primary producers like plankton, plants, trees etc.) are the ones that thrive in an aerobic environment . Overall, high $pO_2$ appears to be an outcome of (large quantities of) life during the Phanerozoic. The presence of high $pO_2$ levels additionally enhances biomass production through the formation of ozone, which shields life from the detrimental effects of shortwave radiation and enables further oxygen and biomass generation[44].

The environmental parameter $pO_2$ indicates a strong negative correlation with biodiversity before 300 Ma ago and a positive correlation from 300 to 0 Ma ago (Table 1). The inverse relationship between $pO_2$ and biodiversity during the earlier time period can be attributed to elevated $CO_2$ levels, which simultaneously functioned as an oxygen sink through decaying biomass and exerted direct adverse effects on organisms, leading to conditions such as hypercapnia in the early Phanerozoic[45]. This correlation may have increased from negative to moderately positive after abundant and highly diverse plant life spread onto the continents and likely facilitated an increase in continental weathering rates, which in turn positively impacted the delivery of nutrients into the marine ecosystem and their diversification[46]. The data suggest that photosynthetic productivity, biodiversity and oxygen levels were enhanced between ~ 360–250 Ma ago, right before and during the time of the early existence of the supercontinent Pangea. We suspected this period to be marked by contrary trends, as the formations of supercontinents lead to large continental deserts and decreased RR due to the absence of proximate water bodies (oceans). However, deserts nowadays are commonly characterized by their highly oxygenated groundwater environments[47]. Alternatively, sufficient land vegetation could have initially acted as an $O_2$ source. Another possibility is that oceanic plankton and other communities of marine primary producers sustained in the vast oceans could have been its source. Nonetheless, sufficiently high oxygen levels during this time span seemed to compensate for decreased relative humidities and temperatures, characterizing an even more thriving biosphere compared to today (see Fig. 3 and Table 2).

Runoff rates are strongly positively correlated with biodiversity throughout the Phanerozoic, especially in combination with higher $T_{glob}$ values. This is in line with the fact that higher RRs, and thus higher continental weathering, provide a large input of macronutrients into the marine environment which are essential for biological diversification. The utilized data suggests that the late Cretaceous to early/mid Tertiary (72–42 Ma ago), the early Carboniferous (355–351 Ma) and the early to mid Silurian (419–415 Ma) were marked by both elevated RRs and $T_{glob}$ compared to their Phanerozoic mean values. These periods were accompanied by $T_{glob}$ and RRs of up to 13% and 35% higher than their mean, respectively, and can be considered as characteristic of a thriving biosphere (Table 2). Contrastingly, periods with insufficient RRs such as the mid to late Jurassic (152 – 124 Ma), the early to mid Jurassic (204 –160 Ma) and the early to mid Triassic (244 –226 Ma) saw some of the lowest biomass and biodiversity (Table 2 and Fig. 3). Some studies[48] indicate that the age and extent of environments with tropical climate throughout Earth's history, rather than the amount of RR alone, provide enhanced opportunities for the diversification of genera. A positive correlation of RR with biomass, which was predominantly composed of photosynthetic organisms throughout the Phanerozoic, is observed in the latter part of the Phanerozoic. This can be explained by the expansion of (plant) life from the oceans onto land towards the second part of the Phanerozoic[49]. However, it is unclear why the correlation is negative for the earlier time period.

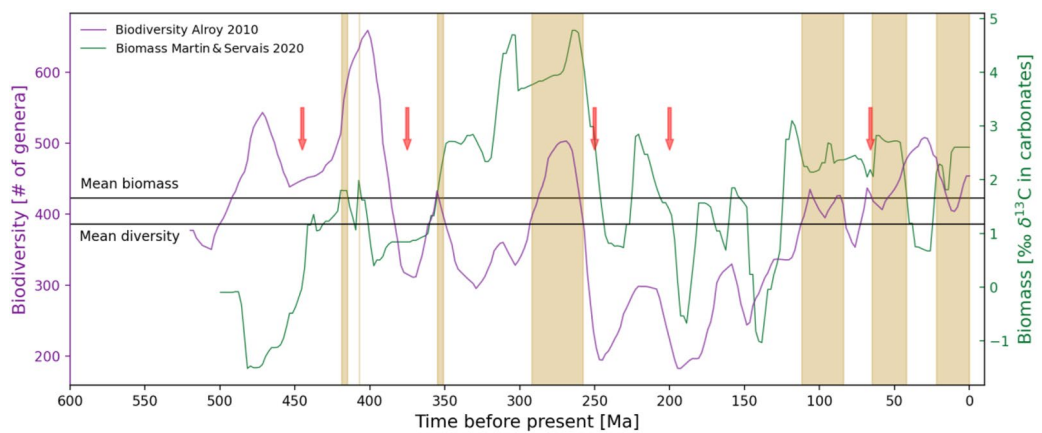

**Figure 3.** Varying biomass and biodiversity values throughout the Phanerozoic. Vertically highlighted are time periods of a more thriving global biosphere relative to the Phanerozoic mean values, i.e. periods during which both the biomass production as well as the biological diversity were higher than their Phanerozoic means. The following mean values were calculated: biomass = 1.65 [‰$\delta^{13}$C in carbonates] for 500–0 Ma; biodiversity = 386 [number of genera] for 520–0 Ma. The red arrows represent the big five mass extinction events. A numerical synthesis of the average values for biomass, biodiversity as well as the accompanying environmental parameters (oxygen partial pressures, global average surface temperatures, runoff rates and atmospheric carbon dioxide) during periods of a more/less thriving biosphere compared to the Phanerozoic mean are presented in Table 2.

| | Periods | Time [Ma] | $pO_2$ | $T_{glob}$ | RR | $CO_2$ | Biomass | Biodiversity |
|---|---|---|---|---|---|---|---|---|
| More habitable biosphere | Mid/late tertiary to quaternary | 0–22 | ↑2.0% | ↓18.6% | ↓14.7% | ↓80.4% | ↑43.0% | ↑11.0% |
| | Late cretaceous to early/mid tertiary | 42–65 | ↑11.4% | ↓15.2% | ↑0.7% | ↓59.9% | ↑54.9% | ↑12.0% |
| | Early to mid cretaceous | 84–112 | ↑19.4% | ↓19.8% | ↓0.8% | ↓16.6% | ↑40.0% | ↑7.0% |
| | Late carboniferous to late Permian | **258–292 | ↑40.6% | ↓21.6% | ↓7.1% | ↓24.9% | ↑149.7% | ↑20.2% |
| | Early carboniferous | 351–355 | ↑16.3% | ↓6.6% | ↑8.3% | ↓18.8% | ↑34.2% | ↑5.5% |
| | Late Silurian | 407–407 | ↑3.4% | ↓13.6% | ↑32.0% | ↑52.2% | ↑19.8% | ↑64.0% |
| | Early to mid Silurian | 415–419 | ↓0.2% | ↑2.3% | ↑35.6% | ↑58.3% | ↑8.3% | ↑48.3% |
| Less habitable biosphere | Mid to late Jurassic | 124–152 | ↑3.6% | ↓1.7% | ↓15.0% | ↓35.3% | ↓87.2% | ↓22.4% |
| | Early to mid Jurassic | *160–204 | ↑4.8% | ↑7.4% | ↓24.1% | ↓78.7% | ↓49.6% | ↓39.2% |
| | Early to mid Triassic | *226–252 | ↑18.7% | ↑30.2% | ↓15.9% | ↓17.7% | ↓23.1% | ↓41.1% |
| | Mid Devonian to early carboniferous | *361–383 | ↓11.6% | ↑8.8% | ↑13.1% | ↓32.7% | ↓46.6% | ↓13.5% |
| | ***Mid Ordovician to early Silurian | *441–455 | ↓40.7% | ↓37.5% | ↑21.3% | ↑58.3% | ↓100.8% | ↑15.3% |

**Table 2.** Time periods of a more/less habitable global biosphere relative to the Phanerozoic mean values, i.e. periods during which both the biomass production as well as the biological diversity were higher/lower than their Phanerozoic means. The accompanying environmental parameters oxygen partial pressures ($pO_2$), global average surface temperatures ($T_{glob}$), runoff rates (RR) and atmospheric carbon dioxide ($CO_2$) are also listed. The following Phanerozoic mean values were calculated: biomass = 1.65 [‰$\delta^{13}$C in carbonates]; biodiversity = 386 [number of genera]; $pO_2$ = 21.2 [%]; $T_{glob}$ = 20.2 [°C]; RR = 48.5 [$10^3$km$^3$/year] and $CO_2$ = 2724.6 ppm. Up/down arrows indicate the percentage increase/decrease of the parameter for the given time span relative to its Phanerozoic mean. For the causality/directionality of the environmental and biological parameters, see Table 1. *Time periods during which a mass extinction event occurred. **Time period during which biomass and biodiversity both surpassed their present day values. ***Period manually included by the authors as it contains a mass extinction event. The period exhibits increased biodiversity despite a mass extinction event due to constraining the comparison to the Phanerozoic biodiversity average.

An intriguing result is the negative correlation of $T_{glob}$ with biodiversity, especially in the most recent 300 Ma. This indicates that raised surface $T_{glob}$, especially in combination with low RRs (e.g. the transitional period between the Triassic and Jurassic ~ 240–160 Ma ago; see Table 2), are detrimental to planetary biodiversity. Rapid changes in $T_{glob}$, such as the global cooling 460–440 Ma ago, can lead to massive loss of biodiversity like during the Ordovician–Silurian extinction (Table 2, Fig. 5 ). Furthermore, a concomitant relationship between $T_{glob}$ and $CO_2$ can be seen in the Permo–Carboniferous (340–260 Ma ago) and the late Cenozoic (35–0 Ma ago) when $CO_2$ levels reached one of two global lows and the Earth experienced the two longest glaciation periods in the Phanerozoic[6]. Moreover, the rapid increase of $T_{glob}$ at the end of the Permian around 252 Ma ago was hypothesized to most likely have been triggered, at least in part, by massive and predominantly volcanic emissions of $CO_2$, which led to one of the largest mass extinction events in Earth's history[7]. Similar negative correlations between $T_{glob}$ and biodiversity, especially for marine families, has previously been established[3,5].

In regards to global climate, all parameters (oxygen content, runoff rate, carbon dioxide levels and temperature) are critical. A decrease in oxygen (and consequently ozone) concentrations decreases the far- and mid-ultraviolet (MUV, FUV) absorption in the atmosphere. This allows more shortwave solar radiation to reach Earth's surface which can have a direct deleterious impact on terrestrial organisms[50,51]. A decrease in RR will decrease the input of macronutrients into coastal marine environments, having detrimental effects on biodiversity and leading to a degradation of ecosystems[21–23]. As our results corroborate, a large and rapid increase of global temperatures is negatively correlated to biodiversity. This is also the reason that anthropogenic emissions of greenhouse gasses nowadays lead to a significant portion of terrestrial and marine species and ecosystems being lost[52–54]. Some authors even predict that a temperature increase at present rates could lead to an extinction event similar to the big five mass extinctions of the Phanerozoic[5].

Beyond Earth, these results can bring us closer to addressing which environmental factors on other planetary bodies may influence quantitative measures of habitability to guide our search for life. Measuring atmospheric biosignatures of terrestrial exoplanets and interpreting atmospheric features in terms of such chemical disequilibria, however, are non-trivial tasks[50,55–57]. The results of this work show that, throughout the Phanerozoic, the main environmental attributes that contributed to the development of a terrestrial floral biosphere include higher oxygen levels and increased global runoff, which can be used to infer the atmospheric humidity or aridity of certain geological time periods, as well as decreased $CO_2$ levels compared to their Phanerozoic mean values. Moderate global average surface temperatures (relative to the Phanerozoic mean) also characterized a habitable biosphere, as long as no large and rapid changes occurred[5]. The amount of $O_2$ characteristic of a thriving terrestrial floral biosphere can be inferred from the planetary spectral features at 0.69, 0.76, 0.77 and 1.27 μm[58]. Its atmospheric proxy $O_3$ can also be used, as it exhibits more dominant features compared to $O_2$ at ~ 0.25, 0.4–0.85 and 9.6 μm[59]. $CO_2$ levels, moreover, can be inferred from the dominant 4.3 and 15 μm features in the (exo)planetary spectra. As we enter the window of technical feasibility, it is increasingly sensible to interpret the simultaneous detectability of spectral features as well as their magnitudes, which are proportional to chemical abundances, as a means to infer the composition of a planetary atmosphere and elucidate whether a biological component may be present.

## Conclusion

Our study emphasizes that Earth experienced periods during which habitability parameters varied strongly throughout the Phanerozoic. A more thriving biosphere (relative to the Phanerozoic mean of biomass and biodiversity) was characterized by higher oxygen levels and runoff rates (which can be used to infer the atmospheric humidity or aridity of a period), as well as moderate global average surface temperatures, as long as no large or rapid temperature excursions occurred. During part of the Permian period, the highest estimated oxygen levels to date seemed to compensate for the decreased runoff rates and temperatures, providing an even more thriving environment compared to today. Contrastingly, a less thriving biosphere was marked by low runoff rates and large positive and negative excursions in temperatures as well as significantly decreased carbon dioxide and/or oxygen levels. We found a direct correlation of oxygen content to a highly thriving biosphere and show that a high oxygen content is diagnostic of increased biomass production in the last 300 Ma when plants dominated the continents. This spreading and abundance of plant life on land also likely facilitated an increase in continental weathering rates, and thus also marine ecosystem diversification through the delivery of nutrients into the oceans. Interestingly, one distinct anomaly seems to be the last 22 Ma exhibiting an increased habitability, despite having both low runoff rates, temperatures and carbon dioxide contents relative to the Phanerozoic mean. Beyond its implications for insights on Earth's varying habitability parameters affecting Earth's climate and biosphere, the results will also be very valuable when assessing whether extrasolar planets might be habitable or even possess a biosphere based on their remotely measured environmental parameters[60–64].

## Methods

**Cross-correlation.** We performed a cross-correlation analysis to quantify the similarities between the environmental parameters (i.e. $pO_2$, $T_{glob}$, RR and $CO_2$) and the biotic variables (i.e. primary productivity as a proxy for biomass, and biodiversity) as well as the correlations between the environmental parameters themselves. The analysis determines the degree of similarity of two time-series with respect to a temporal lag. It provides a correlation value of the two time series at a temporal lag of 0, as well as determines at which temporal lag (μ*) the time series exhibit a maximum correlation. This allowed us to make inferences about potentially causal relationships between parameter-pairs, as the analysis revealed which time-series preceded the other. Caution is required when interpreting the causality of the cross-correlations, as this is dependent on the temporal range of the corresponding errors.

The cross-correlation analysis was conducted in *Python* with the *correlate* function from the *scipy.signal* module. To prepare the data, for each time series we first subtracted the corresponding mean value and normalized it to the maximum (absolute) value. This way each time series fluctuated around 0 and had the same normalized units. We then performed the cross-correlation analysis for each pair of normalized time-series, whose lag-dependent cross-correlation functions (ccf) could be intercompared. We capped our analysis to a ± 25 Ma lag range because ever larger temporal ranges are not expected to be as relevant in a biogeochemical context. The positive and negative errors with respect to μ* were calculated by determining the positive/negative temporal lags at which 34.15% of the total area between the ccf and the horizontal zero-line is enclosed. The analysis was repeated by splitting the entire temporal range at 300 Ma ago to study the spread of ocean life onto land, at which point a change in the correlation between the surface and the biosphere can be expected[49].

**Data.** *Temperature.* Quantitative estimation of Earth's global temperatures throughout its various epochs requires a multifaceted approach, which makes use of isotopic proxy as well as geological and lithological data. Previously, detailed and robust $T_{glob}$ curves have been calculated by mapping the lithography and distribution of rock types onto continental maps of Earth's Phanerozoic eon in order to visualize the spread of the past Köppen belts[35]. The corresponding area coverage of each belt throughout the Phanerozoic has been calculated and, using the mean average temperatures of the modern Köppen belts, the area-weighted average temperatures of each of the climatic belts for the entire Phanerozoic was determined. Since this approach is biased towards the temperatures of the modern Köppen belts, the temperature-sensitive $^{18}O/^{16}O$ ratios from carbonate and phosphatic fossils in the tropics were used in order to correct for this artifact.

In this work we took $T_{glob}$ data provided by Scotese et al.[35] for temperature predictions. Even though this temperature estimation is only one amongst the many attempts to predict Phanerozoic temperatures and relies heavily on modern Köppen belt calculations, their analysis is a multidisciplinary composition of a solid record of geological data and two different independent isotope data sets. Their temperature curve is the product of the collective effort of many earth scientists who revised and refined existing isotopic data.

*Oxygen.* Measuring atmospheric oxygen contents throughout Earth's natural history is still one of the most challenging undertakings, given the notorious lack of direct measurements and the conflicting estimates of atmospheric oxygen concentrations. Previously, atmospheric oxygen concentrations were calculated based on the charcoal record[37]. Inertinite, i.e. fossilized charcoal or oxidized organic material, is the byproduct of wildfires and therefore an indicator of the global atmospheric oxygen content. The lack of fossil charcoal records prior to ~ 400 Ma ago is not necessarily an indicator of a low oxygen environment but might rather point to the lack of wooden plants which combust to produce charcoal[37]. Precise predictions of atmospheric $O_2$ concentrations increasingly also rely on biogeochemical computational models[38,39]. Such models implement state-of-the-art data, such as the sulfur- ($^{34}S/^{32}S$, i.e. $\delta^{34}S$) and carbon- ($^{13}C/^{12}C$, i.e. $\delta^{13}C$) isotope ratios, in order to estimate $O_2$ abundances. *GEOCARBSULFOR*[39] is one of the most recent models which builds upon the original *GEOCARBSULF* model[38] to account for new and improved carbon and sulfur isotope data[65,66]. To predict oxygen concentrations, we calculated the mean of the charcoal record[37], the *GEOCARBSULF*[38] and *GEOCARBSULFOR*[39] model datasets (see Fig. 4).

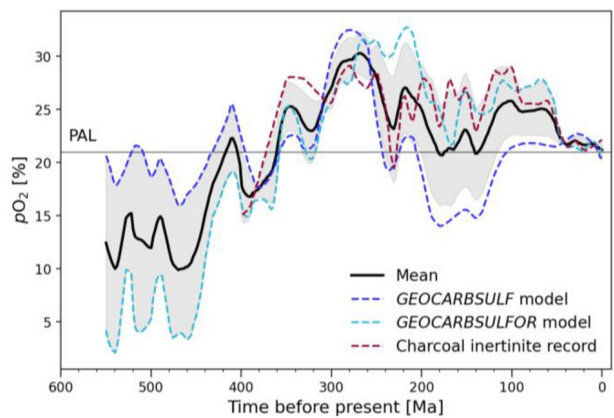

**Figure 4.** Comparison of $pO_2$ data used in this work. They include the charcoal record[37], the *GEOCARBSULFOR* model[39] and the original *GEOCARBSULF* model[38]. The black curve is the calculated mean of all three datasets. The shaded region represents the calculated standard deviation from the mean. PAL = Present Atmospheric Level.

*Global runoff rates.* Explicit data measurements of relative humidity throughout the Phanerozoic are severely lacking but are usually inferred indirectly through various proxies, including the latitudinally and temporally varying distribution of land and water, as well as isotope ratios such as $^{87}Sr/^{86}Sr$, which serve as a quasi water–rock interaction fingerprint[20,27,36,67,68]. An increase in isotope ratio values is a qualitative indicator of increased weathering and runoff rates, since these processes are associated with the input of the heavier isotope $^{87}Sr$ to the oceans[69]. Alternatively, taking into account Earth's varying continental configurations, a projection of modern day latitudinal rainfall, runoff and evaporation rates backwards in time can be used in order to calculate a Phanerozoic runoff curve[27]. These global runoff rates can serve to infer the humidity/aridity of a geological time period, as they are influenced by the Earth's hydrological cycle which is furthermore affected by the continental configurations[20,26–30]. In this work we took values for RR based on the global runoff rates[27].

*Carbon dioxide.* The natural cycle of carbon dioxide in Earth's atmosphere is intimately connected to life on Earth and plays an important role as one of our planet's primary greenhouse gases[15,70,71]. Carbon is necessary for the formation of complex molecules essential for life such as DNA[14]. While most carbon is deposited in rocks and sediments, it can also be found in Earth's atmosphere in the form of carbon dioxide, as well as in oceans and in the biosphere. In general, carbon sinks from the atmosphere, ocean and biosphere include organic carbon burial and the precipitation of inorganic carbonate minerals within marine sediments[71,72]. Carbon is released back into the oceans, atmosphere and biosphere via direct injections from the mantle such as volcanic eruptions and via oxidative/carbonate weathering and degassing of buried carbon[40,71,73]. Many attempts have been made to reconstruct the carbon dioxide reservoir with biogeochemical models, aiming at including all relevant sources and sinks over the Phanerozoic Eon[6,40,71,73–75].

Here we took $CO_2$ data provided by Mills et al.[40] for Phanerozoic carbon dioxide reconstructions. Their $CO_2$ curve is calculated with the Spatial Continuous Integration (*SCION*) model which bridges the spatial climate modeling procedure in *GEOCLIM*[75] and the long-term biogeochemical processes in the *COPSE* model[76].

*Biomass.* The primary productivity of a biological system is the rate of production of biomass (i.e. size + abundance) per unit area or volume by autotrophic photosynthetic organisms (e.g. plants), the primary producers[20]. Primary productivity thus serves as a proxy for biomass throughout the Phanerozoic. Photosynthetic organisms make up ~ 82% of the total biomass on Earth today, and are thus its dominant source[41]. Autotrophic photosynthetic organisms tend to use the lighter carbon isotope in $CO_2$ during photosynthesis (e.g. $^{12}C$ rather than $^{13}C$), thus effectively locking it out of the atmosphere. This shifts the atmospheric $^{13}C/^{12}C$ ratio to higher values. Thus, a higher isotopic ratio in sedimentary (i.e. inorganic) carbonates, which reflect the atmospheric $^{13}C/^{12}C$ ratio, is also associated with increased primary i.e. photosynthetic productivity[20,77]. In this work we took values for primary productivity[36] as a proxy for biomass based on previous work[67].

*Biodiversity.* Most biodiversity curves in the literature are based on data from marine fossils, because such organisms inhabited locations nearest to where sediments usually are deposited[78]. Generally, biodiversity curves are highly influenced by the method used to count taxa in a given time interval which can lead to a bias towards the record of extant data, which is much more numerous and better preserved than data from the more distant past. Alroy[34] used a novel method of standardized sampling and counting of taxa from the fossil record named shareholder quorum subsampling(SQS), in order to correct for possible biases. For that analysis, fossil data from the Paleobiology Database (PaleoDB: http://paleodb.org) downloaded in the year 2010 were used.

We focused on data from Cambrian trilobites, Paleozoic brachiopods, and modern gastropods[79]. The diversity pattern of these three so-called "evolutionary faunas" (EF) expanded and contracted in a similar fashion throughout the different phases of the Phanerozoic[79–82] (see *Gastropoda*, *Brachiopoda* and *Trilobita* curves in Fig. 5a). For our analysis, we utilized the sum of the corrected-for-bias values of the three EF based on Alroy[34]. We did this by summing up the corrected *Gastropoda*, *Brachiopoda* and *Trilobita* curves in Fig. 5b for each time-stamp in the past 520 Ma (see light blue curve in Fig. 5b).

A summary of all the input data used for this work is provided in Table 3.

## Data availability

All data generated or analyzed during this study are included in this published article and their sources are cited where applicable.

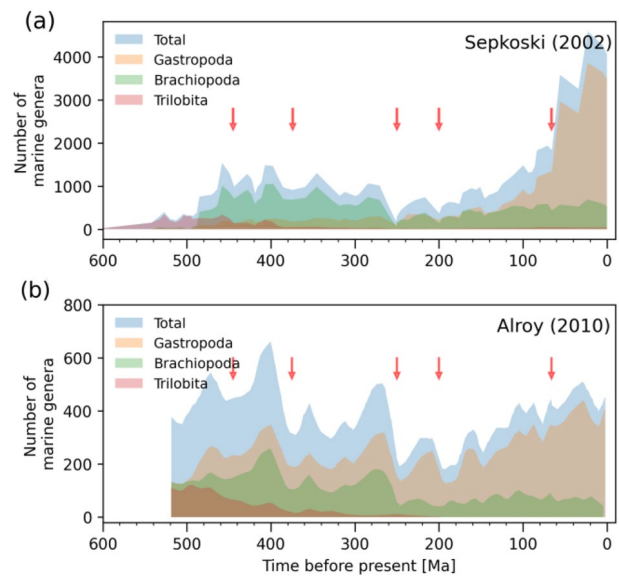

**Figure 5.** Phanerozoic biodiversity curves. (**a**) Total Phanerozoic diversity based on raw data[82] divided into its three main Evolutionary Faunas (EF), i.e. the Cambrian trilobites, the Paleozoic brachiopods, and the modern gastropods. These are three distinct types of fauna whose diversity pattern expanded and contracted in a similar fashion throughout the different phases of the Phanerozoic. Each EF includes specific taxonomic classes that were most abundant during (but not limited to) a certain period in Earth's history. Furthermore, the EFs successively replaced one another as the principal group throughout the Phanerozoic. The red arrows represent the Big Five mass extinction events, including the Ordovician–Silurian extinction (~ 445 Ma ago), Devonian extinction (~ 375 Ma ago), Permian–Triassic extinction (~ 250 Ma ago), Triassic-Jurassic extinction (~ 200 Ma ago) and the Cretaceous-tertiary extinction (~ 66 Ma ago)[81]. (**b**) Corrected biodiversity curve using the shareholder quorum subsampling (SQS) sample standardization method[34].

<200b>www.nature.com/scientificreports/

| Parameter | Method | Proxy | Available range [Ma] | Source |
|---|---|---|---|---|
| $pO_2$ | Charcoal inertinite record | – | 400–0 | 37 |
| " " | GEOCARBSULFOR model | Carbon & sulfur cycles | 550–0 | 39 |
| " " | GEOCARBSULF model | Carbon & sulfur cycles | 550–0 | 38 |
| $T_{glob}$ | Köppen belts + $\delta^{18}O$ carbonate/phosphatic fossils | – | 550–0 | 35 |
| RR | Projection of hydrologic cycle | Global runoff rates | 560–0 | 42 |
| Biomass | $\delta^{13}C$ inorganic (carbonate) fossil data | Primary productivity | 500–0 | 36 |
| Biodiversity | SQS sample standardization | – | 520–0 | 34 |
| $CO_2$ | SCION model | – | 550–0 | 40 |

**Table 3.** Summary of data used in this work. Atmospheric oxygen partial pressure ($pO_2$), global average surface temperature ($T_{glob}$), runoff rate (RR), atmospheric carbon dioxide ($CO_2$) as well as biomass and biodiversity data were compiled from the corresponding references listed in the last column.

### Acknowledgements
The first author would like to thank the scholarship organization *Studienstiftung des deutschen Volkes* without which this work would not have been possible. The authors also thank Jan-Peter Duda, Christof Sager and Jacob Heinz for their valuable comments on a draft version of this manuscript.

### Author contributions
Conceptualization: D.S.M., Methodology: IV, D.S.M., R.H., Investigation: IV, Visualization: IV, Formal analysis: IV, R.H., Funding acquisition: IV, Supervision: D.S.M., R.H., Writing—original draft: IV.

### Funding
Open Access funding enabled and organized by Projekt DEAL. *Studienstiftung des deutschen Volkes* doctoral scholarship (IV) PLATO Data Center grant 50OO1501(RH).

### Competing interests
The authors declare no competing interests.

### Additional information
**Supplementary Information** The online version contains supplementary material available at https://doi.org/10.1038/s41598-023-39716-z.

**Correspondence** and requests for materials should be addressed to I.V.

**Reprints and permissions information** is available at www.nature.com/reprints.

**Publisher's note** Springer Nature remains neutral with regard to jurisdictional claims in published maps and institutional affiliations.